\documentstyle[aps,prl,twocolumn,epsf,epsfig,rotate]{revtex}
\newcommand{\rr}{{\bf r}}

\newcommand{\qq}{{\bf q}}

\newcommand{\GG}{{\bf G}}

\newcommand{\ah}{{\hat a}}

\newcommand{\ahd}{{\hat a^\dagger}}

\begin{document}
\twocolumn[\hsize\textwidth\columnwidth\hsize\csname
@twocolumnfalse\endcsname

\title{Quantum Melting and Absence of Bose-Einstein Condensation 
in Two-Dimensional Vortex Matter} 

\draft
\author{Jairo Sinova$^{(a)}$, C.B. Hanna$^{(b)}$, and A.H. MacDonald$^{(a)}$}
\address{$^{(a)}$ Department of Physics,
University of Texas, Austin, Texas 78712-1081\\
$^{(b)}$ Department of Physics,
Boise State University, Boise, Idaho 83725-1570}
\date{\today}
\maketitle

\begin{abstract}
We demonstrate that quantum fluctuations suppress Bose-Einstein condensation
of quasi-two-dimensional bosons in a rapidly-rotating trap.
Our conclusions rest in part on the derivation of an exact expression for
the boson action in terms of vortex position coordinates,
and in part on a solution of the weakly-interacting-boson Bogoliubov equations,
which simplify in the rapid-rotation limit.
We obtain analytic expressions for the collective-excitation dispersion,
which is quadratic rather than linear.
Our estimates for the boson filling factor at which the
vortex lattice melts are consistent with recent exact-diagonalization calculations.
\end{abstract}
\pacs{PACS: 03.75.Fi, 05.30.Jp, 73.43.Cd, 73.43.Nq}

\vskip2pc]

Recent experiments \cite{Ketterle,Madison} 
in rapidly-rotating Bose-condensed alkali gases \cite{general} 
have established the occurrence of large vortex arrays.  These 
observations  have raised a number of fundamental questions about 
boson vortex matter that are suggested by loose analogies with the properties 
of type-II superconductors in a magnetic field, and by the lore of
the boson quantum Hall effect \cite{Ho}.
The simplicity of dilute alkali gases, 
and the arsenal of techniques that have been developed 
to manipulate them experimentally, make these systems ideal for the study of
condensed-matter systems containing vortices.
The proportionality between rotation frequency
and rotating-frame effective magnetic field suggests \cite{Fetter}
that there will be a maximum rotation frequency, $\Omega_{c2}$, beyond which 
vortex-lattice states cannot occur, even at zero temperature.  
This expectation is consistent \cite{Wilkin2000} with our understanding
from quantum-Hall studies that quasi-two-dimensional (2D) charged-boson
systems in strong magnetic fields form strongly-correlated fluid ground
states that do not break translational symmetry.
In theoretical studies of rapidly-rotating boson systems,
exact-diagonalization studies tend to suggest that the bosons form
fluid states, while Thomas-Fermi and mean-field studies predict  
vortex-lattice states.
Very recently, it has been suggested \cite{Cooper,Liu}, on the basis of
exact-diagonalization studies using periodic boundary conditions, 
that the vortex-lattice state of quasi-2D bosons melts due to quantum
fluctuations when the boson filling factor $\nu$, the ratio of boson
density to vortex density, is smaller than $\sim 6$.  

In this Letter, we present a theoretical study of quantum fluctuations in 
rotating Bose-Einstein condensates that is based partly on a long-wavelength
effective Lagrangian derived from the microscopic action, 
and partly on microscopic Bogoliubov-approximation calculations in the 
rapid-rotation limit.  We find that the collective excitation energy of 
quasi-2D rotating bosons has a wave-vector ($q$) dependence that is
quadratic at small $q$, in contrast to the linear behavior ordinarily
characteristic of interacting boson systems.  
The quadratic energy dispersion has fundamental consequences for 
the nature of the interacting-boson ground state:
for any $\nu$, the integral for the fraction of particles outside the 
condensate diverges logarithmically at small $q$. Thus,
{\em Bose-Einstein condensation does not occur even at zero temperature
in the thermodynamic limit.} 
Fluctuations need not cause the vortex lattice to melt, however,
and our results for the melting filling factor are consistent with
recent exact-diagonalization\cite{Cooper,Liu} estimates. 

A system of interacting bosons of mass $M$ 
in a cylindrical trap (with radial and axial trap frequencies 
$\Omega_r$ and $\Omega_z$) that is rotating with angular velocity
$\Omega {\bf \hat{z}}$ is well-described by the rotating-frame
Hamiltonian\cite{general,Cooper}
\begin{eqnarray}
\cal{H}&=&\sum_{i=1}^N \left\{ \frac{({\bf p}_i -
M\Omega {\bf \hat{z}}\times \rr_i)^2}{2M}
+\frac{M}{2} \left[(\Omega_r^2-\Omega^2)(x_i^2+y_i^2) \right. \right.
\nonumber\\&& \left. \left.
+ \Omega_z^2z_i^2 \right] \right\} +\sum_{i<j=1}^N V(\rr_i-\rr_j) ,
\label{hamiltonian}
\end{eqnarray}
where $V(\rr)$=$g\delta(\rr)$ is a hard-core interaction potential
of strength $g$.
This Hamiltonian is equivalent to that of a system of charge-$Q$ bosons
with weakened radial confinement, under the influence of a 
magnetic field ${\bf B}=(2M\Omega/Q){\bf\hat{z}}$.
The theory simplifies in the quasi-2D, rapid-rotation,
weak-interaction limit, in which the inequalities
$\Omega \gg \sqrt{\Omega_r^2-\Omega^2} > 0$, 
$ng < \Omega_z$ and $ng < 2 \Omega$ are satisfied.
(Here $n$ is the boson number density.)  All the calculations discussed
here use the rotating-frame Hamiltonian and assume the rapid-rotation limit
in which only single-particle states in the lowest Landau level
(LLL, those states that have the minimum quantized kinetic energy)
are relevant.
(For weakly-interacting bosons, filling factors $\nu$ larger than one
are compatible with the LLL constraint.)
However, as we explain later, we believe that our most important 
conclusions are more general.   
In the quasi-2D weak-interaction limit, properties of the system
depend non-trivially only on the boson filling factor 
\begin{eqnarray}
\nu={N}/{N_V} \equiv {2\pi \ell^2 N}/{A} \equiv 
({N}/{A})({2\pi\hbar}/{2M\Omega}).
\end{eqnarray}
Here $A$ is the cross-sectional area of the rotating system of $N$ bosons,
$N_V$ is the number of vortices in the system, 
and $\ell$=$\sqrt{\hbar/2M\Omega}$ is the effective magnetic length.
In the following paragraphs, we consider unconfined
($\Omega$$\to$$\Omega_r$)
but interacting 2D bosons in the LLL \cite{noteg2d}.
We discuss quantum fluctuations in the vortex-lattice ground state
by first using a coherent-state path-integral description
in terms of vortex-position coordinates, which makes the physical origin
of our surprising findings apparent.
We then work directly with the boson coordinates by using a
Bogoliubov approximation, which is a simpler approach for
quantative analysis of the weak-fluctuation limit.

The interacting-boson partition function  is
{$Z=\int {\cal D}[\bar{\phi({\rr})}\phi({\rr}) ] \exp(-S[\phi])$},
where the action is
\begin{eqnarray}
S&=&\int_{0}^{\infty} d\tau\int d\rr\left\{
\bar\phi(\partial_\tau-\mu)\phi+\frac{g}{2}
|\phi|^4\right\}\nonumber \\
&\approx& S_{MF}+  
\int_{0}^{\infty} d\tau\int d\rr\left\{
i \delta n \partial_\tau\delta \theta
+\frac{g}{2}(\delta n)^2\right\}.
\label{eq:actionsf}
\end{eqnarray} 
In the approximate Gaussian-fluctuation form of this action that we employ,
the amplitude and phase contributions are separated:
\mbox{$\phi = \sqrt{n} e^{i\theta}$}.
The usual gradient terms are absent in Eq.~(\ref{eq:actionsf})
because of the restriction to the LLL, and we have dropped the constant 
boson kinetic energy, $N\hbar\Omega$.
The imaginary-time-dependent coherent-state label $\phi({\rr},\tau)$
satisfies the LLL constraint, which guarantees that 
it is completely specified by the locations of its zeroes,
up to an overall constant.  In the mean-field ground state, the zeroes of 
$\phi$ are located at the sites of an Abrikosov triangular vortex lattice.
We express the action in terms of fluctuating vortex positions, $u_i$,
by writing $\phi$ as a symmetric-gauge LLL wavefunction\cite{GirvinJach}:
\begin{equation}
\phi[u] = \Phi_u\prod_{i=1}^{N_V} ({z} 
 - z_i-u_i) \exp( -z\bar z /4\ell^2) = \sqrt{n}e^{i\theta} ,
\end{equation}
where $z=x+iy$.  Here the $z_i$ are the triangular-lattice vortex sites,
and the vortices of $\phi[u]$ are located at $z=z_i+u_i$:
i.e., $u_i$ is the displacement of the $i$-th vortex.
It follows from the LLL constraint that the order-parameter phase
and amplitude fluctuations are given to leading order in $u$ by 
\begin{eqnarray}
&& \theta \approx {\rm Im} \sum_{i} u_i \frac{d[\ln(z-z_i)]}{dz_i}
= \sum_i \frac{(y-y_i)u_{xi}-(x-x_i)u_{yi}}{|z-z_i|^2} , \nonumber\\
&&\delta n \approx 
2n_0 \sum_i \frac{(x_i-x)u_{xi}+(y_i-y)u_{yi}}{|z-z_i|^2}\,.\nonumber
\end{eqnarray} 
Fourier-expanding the fluctuation action, we find that at long wavelengths
it assumes the form 
\begin{eqnarray} 
\delta S 
\sim \hspace {-0.06in}\int_{0}^{\infty} \hspace {-0.17in}
d\tau \frac{1}{A} \sum_\qq \left\{
i\frac{\alpha_1}{q^2} \bar u_L\partial_\tau u_T
+\frac{\alpha_2}{q^2} |u_L|^2
+ c_{66}q^2 |u_T|^2 \right\} ,
\nonumber
\end{eqnarray}
where $u_L$=$[q_x u_x(\qq,\tau)+ q_y u_y(\qq,\tau)]/q$ is the
longitudinal vortex displacement,
$u_T$=$(q_x u_y-q_y u_x)/q$ is the transverse vortex displacement,  
$c_{66}$ is a shear modulus that appears here as a phenomenological
coefficient (but see below), $\alpha_1$=$2 n(\GG$=$0)/l^4$,
$\alpha_2$=$2g n^2(\GG$=$0)/\ell^4$, $|\phi_0(z)|^2\equiv\sum_\GG n(\GG)
\exp(i\GG\cdot\rr)$, $|\phi_0(z)|^4\equiv\sum_\GG n^2(\GG)\exp(i\GG\cdot\rr)$,
and $\GG$ denotes the reciprocal-lattice vectors of the triangular lattice
\cite{note_nick}.

A closely-related classical energy functional arises in the theory of 
thermal vortex-line fluctuations in type-II superconductors,
and leads in that case to the loss of long-range coherence
\cite{Moore,Maki,Tesanovic} at finite temperatures.
Here, it follows from our analysis that true off-diagonal long-range order,
and hence Bose-Einstein condensation, is destroyed at zero temperature
by quantum fluctuations for any value of $\nu$,
since $\langle |\theta(\qq)|^2\rangle\propto 1/q^2$,
characteristic of 2D superfluidity with quasi-long-range
(algebraically-decaying) off-diagonal order.
Quantum fluctuations do not necessarily melt the vortex lattice, however, 
since, according to the above action,  $\langle |u(\qq)|^2\rangle$
goes to a constant as $q$$\to$0,
impying that the vortex lattice can still retain true positional
long-range order.
The diverging phase fluctuations of this action are accompanied by 
a gapless collective mode\cite{typeIImode} with quadratic dispersion,
$E_{\rm eff}(q)\approx \sqrt{2\beta_{AB} g c_{66}} (q\ell)^2$,
where $\beta_{AB}$=$n^2(\GG$=$0)/(n(\GG$=$0))^2$=$1.1596$.   
To estimate the filling factor at which the vortex lattice
melts due to quantum fluctuations,
we need to evaluate the shear constant $c_{66}$, which we now extract from 
a fully-microscopic Bogoliubov-approximation calculation.

The rotating-frame Hamiltonian [Eq.~(\ref{hamiltonian})]
in second-quantized form is
\begin{eqnarray}
{\hat{\cal{H}}}&=&\frac{1}{2}\sum_{{X_1}',{X_2}',X_1,X_2}
\langle {X_1}',{X_2}'|V|X_1,X_2\rangle \ahd_{{X_1}'}
\ahd_{{X_2}'}\ah_{X_2}\ah_{X_1} ,
\nonumber
\end{eqnarray}
where we have switched to the Landau gauge for a more convenient
microscopic description of vortex-lattice states \cite{gaugeinvar},
and $\ahd_{X}$ creates 
a Landau-gauge periodic-boundary-condition single-particle state
\[
\phi_{X}(\rr)=\frac{1}{\sqrt{L_y \ell \sqrt{\pi}}}
\sum_{s=-\infty}^\infty e^{i\frac{1}{\ell^2}X_sy}
e^{-\frac{1}{2 \ell^2}(x-X_{s})^2}\,\,,
\]
where $X_{s}= X+s L_x$ with $0<X<L_x$,
and $X L_y/\ell^2$ is an integer multiple of $2 \pi$. 
The interacting-boson partition function may be expressed as a
coherent-state path integral in which the LLL constraint
is imposed explicitly:
$Z=\int {\cal D}[\bar{\psi}_X\psi_X] \exp(-S[\psi])$, where
\begin{eqnarray}
S[\psi]&=& \int_0^\beta d\tau \left[
\sum_X \bar{\psi}_X (\partial_\tau-\mu) \psi_X
+ \frac{1}{2} \sum_{{X_1}',{X_2}',X_1,X_2}\right.\nonumber
\\ \nonumber && \left. 
\langle {X_1}',{X_2}'|V|X_1,X_2\rangle
\bar{\psi}_{{X_1}'}\bar{\psi}_{{X_2}'}{\psi}_{X_2}{\psi}_{X_1} \right]
\end{eqnarray}
is the action.
Minimizing this action leads to the strong-field limit of the
Gross-Pitaevskii (GP) equation,
\begin{equation}
\sum_{{X_2}',X_2,X} \bar{\psi}^0_{{X_2}'}{\psi}^0_{X_2}{\psi}^0_{X}
\langle X',{X_2}'|V|X,X_2\rangle =\mu{\psi}^0_{X'}\,\,.
\label{MFeq}
\end{equation}
A similar equation arises near the upper critical field 
in the Ginzburg-Landau theory of type-II superconductors;
the role played here by the chemical potential ($\mu$) is assumed in that 
case by the temperature, measured from its mean-field-theory critical value.
The global minimum of Eq.~(\ref{MFeq})
is known to occur for the triangular vortex-lattice state,
which has the following form when expressed in terms of
Landau-gauge single-particle states:
$\Psi^0(\rr)=\sum_X \psi^0_X \phi_X(\rr)$,
where
$\psi^0_X=c_\triangle \big[ \sum_n
\delta_{X,n a_\triangle}+i\delta_{X,(2n+1)a_\triangle/2}\big]$,
and $a_{\triangle}^2=4\pi/\sqrt{3}\ell^2$.
The contribution to the filling factor from condensate bosons,
$\nu_0$=$c_\triangle^2$, is related to $\mu$ as described below. 

Expanding the action to second order around the solution of 
the GP equation yields the Bogoliubov approximation for  
interacting bosons.  In the strong-field limit,
\begin{eqnarray}
S&=&S_{MF}\hspace{-.02in}+\hspace{-.05in}\int_0^\beta\hspace{-.1in} d\tau
\hspace{-.02in}\left[
\sum_X \bar{\psi}_X\partial_\tau\psi_X
+\frac{1}{2}\sum_{X',X}\hspace{-.03in}\left\{ \bar{\psi}_{X'}2[E(X',X)\right.\right.\nonumber\\
&-&\mu\delta_{X',X}]{\psi}_{X}\hspace{-.05in}
\left.\left.+\bar{\psi}_{X'}\bar{\psi}_{X}\Lambda(X',X)\hspace{-.02in}
+{\psi}_{X'}{\psi}_{X}\bar{\Lambda}(X',X)\right\}\right],
\nonumber
\end{eqnarray}
where 
\begin{eqnarray}
E(X',X)&=&\sum_{X_1',X_1}\left[ \langle X' X_1'|V\left(|X,X_1\rangle
 + |X_1,X\rangle\right)\right] \bar{\psi}^0_{X_1'}{\psi}^0_{X_1} ,
\nonumber\\
\Lambda(X',X)&=&\sum_{{X_1}',X_1}
\langle X'X|V|{X_1}',X_1\rangle{\psi}^0_{{X_1}'}{\psi}^0_{X_1} .
\nonumber
\end{eqnarray}
For the hard-core model considered here,
the direct and exchange terms in $E(X',X)$ are identical.
The action can be further simplified by using the translational symmetry
of the vortex lattice to transform to a magnetic-Bloch-state representation.
Then
\begin{eqnarray}
S&=&S_0+
\int_0^\beta d\tau {\sum_{\qq\in{\rm BZ}}}^\prime \left[
\bar{\psi}_\qq(\partial_\tau-\mu)\psi_\qq\right.\nonumber\\
&+&\left.\frac{1}{2} (\bar{\psi}_{\qq}\xi_\qq{\psi}_{\qq}
+\bar{\psi}_{-\qq}\xi_\qq\bar{\psi}_{-\qq}
+\lambda_\qq\bar{\psi}_{\qq}\bar{\psi}_{-\qq}
+\bar{\lambda}_\qq{\psi}_{-\qq}{\psi}_{\qq})\right] ,
\label{eq:action}
\end{eqnarray}
where the wave-vector is summed over the triangular-lattice Brillouin zone, 
the prime indicates that $q$=0 is excluded from the sum, and 
$\xi_\qq=2\epsilon_\triangle(\qq)-\mu$, with
\begin{eqnarray}
\epsilon_\triangle(\qq)&=&\frac{g\nu_0}{2\pi \ell^2}
\sum_{\GG} e^{-\frac{\ell^2}{2}|\qq-\GG|^2},
\label{epsilon}
\end{eqnarray}
$\mu=\epsilon_\triangle(0) = \beta_{AB} g\nu_0/(2\pi\ell^2) $, and
\begin{eqnarray}
\lambda_\qq&=&
b \sum_{k,n} e^{-iq_x(k+n)a_\triangle}
\left[ e^{-\frac{1}{2\ell^2}(q_y\ell^2+ka_\triangle)^2}
e^{-\frac{1}{2\ell^2}(q_y\ell^2+na_\triangle)^2}\right.\nonumber\\
&-&e^{-\frac{1}{2\ell^2}(q_y\ell^2+(2k+1)a_\triangle/2)^2}
e^{-\frac{1}{2\ell^2}(q_y\ell^2+(2n-1)a_\triangle/2)^2}\nonumber\\
&+& 2e^{-iq_xa_\triangle/2}
\left. e^{-\frac{1}{2\ell^2}(q_y\ell^2+(2k+1)a_\triangle/2)^2}
e^{-\frac{1}{2\ell^2}(q_y\ell^2+na_\triangle)^2} \right] ,
\nonumber
\end{eqnarray}
where $b$=$({g\nu_0}/{4\pi \ell^2})\sqrt{2/\sqrt{3}}$.
The quadratic action in Eq.~(\ref{eq:action}) has the familiar form
that is diagonalized by a Bogoliubov transformation \cite{F&W}.
It follows that the elementary collective-excitation energy is given by 
\mbox{$E_\qq=\sqrt{\xi_\qq^2 -|\lambda_\qq|^2}$}, and that
the filling factor for bosons outside the condensate is 
\begin{eqnarray}
\nu-\nu_0 = \frac{\ell^2}{4\pi}\int_{BZ} d\qq ~ (\xi_\qq/E_\qq - 1) .
\label{nu_minus_nu0}
\end{eqnarray}
We find, in agreement with our long-wavelength analysis, that 
the collective-excitation energy has a $q^2$ dependence at small $q$
($E_q$$\sim$$0.0836 g\nu_0 q^2$ for $q\ell$$\to$0).
As a consequence, the density of particles outside the condensate diverges
in the thermodynamic limit, consistent with divergent condensate 
phase fluctuations.  As in the effective-action theory,
the vortex-lattice-state density-wave order parameter $\rho(\GG)$  
lacks divergent fluctuations.
Comparing expressions for the collective-excitation energy as
$q$$\to$$0$ yields
$c_{66}\sim(0.0836)^2g \nu_0^2/(2\beta_{AB} \ell^4)$ and, 
upon applying a Debye-like model to extend the effective theory to shorter
wavelengths, $\sqrt{\langle |u_i|^2\rangle} \sim 1.15 \ell/\sqrt{\nu}$.
It then follows from the Lindemann criterion \cite{lindemann} that the
vortex lattice melts due to quantum fluctuations for $\nu$ smaller than
$\sim 8$, consistent with exact-diagonalization estimates \cite{Cooper}.

\begin{figure}
\epsfxsize=2.90in
\centerline{\epsffile{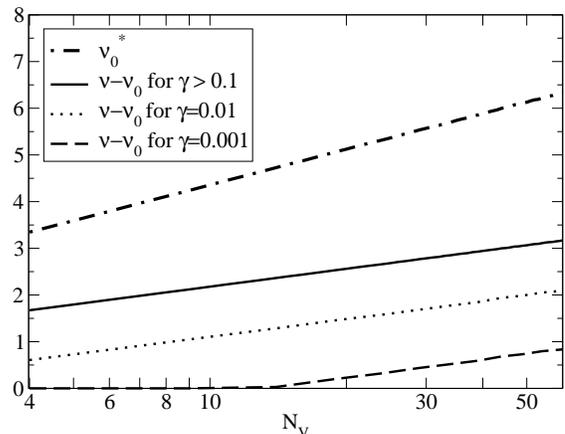}}
\caption{Filling-fraction contribution from bosons outside the condensate,
$\nu$-$\nu_0$, versus $N_V$.  The effect of Landau-level mixing
is estimated by including only collective fluctuations
with energy smaller than the Landau-level separation $2\hbar\Omega$.
The three curves are for
$\gamma\equiv2\ell^2\hbar \Omega/(g\nu_0) >0.1$ (solid line),
$\gamma$=0.01 (dotted line), and $\gamma$=0.001 (dashed line).
The minimum condensate filling fraction $\nu_0^*$, below which
the condensate disappears, is shown as the dash-dotted line.}
\label{numinusnu0}
\end{figure}

The rotation frequency at which Bose-Einstein condensation will no longer 
be evident in realistic finite-size trapped-boson systems
can be estimated using Eq.~(\ref{nu_minus_nu0}).
In the rapid-rotation limit, ($\nu$-$\nu_0$) is independent of the
interaction strength $g$, but depends on $N_V$ through the system-size
infrared cutoff, which we estimate as
$q_{\rm min}$=$1/\sqrt{2\pi N_V \ell^2}$.
The resulting plot of ($\nu$-$\nu_0$) versus system size (i.e., $N_V$)
is shown in Fig.~\ref{numinusnu0}.
Adopting $(\nu-\nu_0) < \nu_0/2$ as a practical criterion for
experimentally-evident Bose-Einstein condensation leads to the
dot-dashed line in Fig.~\ref{numinusnu0}.
The experimental consequences of enhanced quantum phase fluctuations in 
finite-size rotating traps will be addressed at greater length in future work.

Although our discussion up to this point has been restricted to the 
rapid-rotation limit, we believe that our main conclusions are more 
general, because excitations to higher Landau levels are massive 
and become irrelevant \cite{Moore} at low energies.
The effect of quantum fluctuations on the condensate fraction at moderate
rotation rates can be roughly estimated by introducing an ultraviolet cutoff
in the fluctuation integral in Eq.~(\ref{nu_minus_nu0}) at
$E_{q_{\rm max}} = 2 \hbar \Omega$: i.e.,
keeping only fluctuations in which the boson kinetic energy does not increase.
Calculations of ($\nu$-$\nu_0$) for different values of $E_{q_{\rm max}}$
are shown in Fig.~\ref{numinusnu0}.
In the vortex systems studied in recent experiments \cite{Ketterle,Madison},
$N_V$$\sim$$100$ and $\nu_0\sim 10^4-10^5$; for these systems,
$(\nu-\nu_0)/\nu_0\sim 10^{-4}$, by this crude estimate that neglects
fluctuations of the boson kinetic energy.  We conclude that
the strengthening of quantum fluctuations that we predict
for the rapid-rotation limit does not play an important role in these
experiments.
Vortex-fluid states will tend to occur as $\Omega$$\to$$\Omega_r$:
i.e., when the rotating-frame trap frequency vanishes.
At sufficiently small values of $\nu$, we expect these vortex-fluid states
to Wigner-crystallize into lattices of bosons rather than lattices of
vortices.  
Finally, we remark that confinement in trapped-boson systems will
induce a gap in the collective-excitation spectrum, 
which will reduce the importance of quantum fluctuations.

In conclusion, we have shown that the zero-point motion
of the boson vortex lattice in a large rapidly-rotating quasi-2D trap
eliminates Bose-Einstein condensation even at zero temperature,
in the thermodynamic limit.
As a result, the condensate fraction in a finite-size trapped-atom system 
may be relatively small, even when the interactions between bosons are weak.
We predict that the fraction of bosons outside the condensate will grow
logarithmically with the system size, but that the vortex-lattice will
not necessarily be melted by such fluctuations.  
Our estimate for the filling factor below which the lattice
melts due to quantum fluctuations,
$\nu$$\sim$8, is consistent with exact-diagonalization estimates
\cite{Cooper}.
Our estimate of the size ($N_V$) at which Bose-Einstein condensation is lost,
shown in Fig.~\ref{numinusnu0}, could be tested in current traps 
if the boson density could be reduced to reach the low
filling factors at which quantum-fluctuation effects become important.
Our predictions add to the fundamental interest\cite{Ketterle} in 
trapped-boson vortex-matter studies. 

The authors acknowledge helpful discussions with T. Nakajima, J. Fernandez
and N. Read.
This work was supported by the Welch Foundation and the by the National Science
Foundation under grants DMR-0115947 and DMR-9972332.  CBH and AHM are grateful
for the hospitality of St. Francis Xavier University, Antigonish, NS,
where this work was initiated.


\vskip 0.08 in


\begin{references}
\bibitem{noteg2d} The effective interaction in the quasi-2D limit is
$g_{2D}$=$g \delta^{(2)}({\rr}) \int dz \chi^4(z)$,
where $\chi(z)$ is the axial-direction trap wave function.
We have set $g_{2D}$$\to$$g$ in our discussion.
In neglecting confinement, we are assuming that
$\Omega_r^2-\Omega^2$$\ll$$\Omega^2$.
\bibitem{gaugeinvar} The symmetric gauge is not necessarily preferred
when confinement is neglected, and our physical results are gauge-invariant.
\end{references}
\end{document}